# The Model of Gas-Discharge Nonneutral Electron Plasma

N. A. Kervalishvili [1] and G. N. Kervalishvili [2]

[1] Andronikashvili Institute of Physics, Javakhishvili Tbilisi State University,
Tbilisi 0177, Georgia.  <n_kerv@yahoo.com>

[2] Helmholtz Centre Potsdam, GFZ German Research Centre for Geosciences,
Telegrafenberg, 14473 Potsdam, Germany.

**Abstract.** The model of gas-discharge nonneutral electron plasma has been considered, in which the electron density is limited by non-linear processes initiated by diocotron instability and it does not depend on the mechanism of electron transport across the magnetic field. The model describes well the characteristics of electron sheath and the current characteristics of discharge both, in magnetron geometry and in the geometry of inversed magnetron, and it allows us to describe quantitatively the influence of anode misalignment on the discharge electron sheath for the first time. The scope of applicability of the proposed model, as well as its relation with other models of electron sheath is studied.

## I. INTRODUCTION

In nonneutral plasmas the predominant type of interaction of charged particles is electrostatic repulsive force. Therefore, the laboratory nonneutral plasmas can be confined only by a strong magnetic field. For this purpose, the devices with crossed electric and magnetic fields are used. In the simplest case, the discharge device consists of two coaxial cylindrical electrodes located in longitudinal magnetic field. One of the electrodes serves as an anode (external, in case of magnetron or internal – in case of inverted magnetron) and the second one – as a cathode. Along the magnetic field, the discharge space between the cylinders is limited by end electrodes, which are under the cathode potential. In these devices, the nonneutral electron plasma can be easily obtained by the discharge. The parameters of the discharge are such that the ions are not magnetized and leave the discharge gap without collisions, while the electrons are strongly magnetized and are captured by magnetic field. Along the magnetic field the electrons are kept by electrostatic fields. Under these conditions, near the anode surface the sheath of nonneutral electron plasma is formed and the whole discharge voltage falls on it [1-8].

Already the first experimental results (the dependence of electron density and discharge current on pressure, and the thickness of electron sheath) indicated that the main mechanism of electron transport across the magnetic field can be the classical mobility. In this case, from the theory it follows that the discharge current should be proportional to magnetic field. However, the experiment always showed the independence or the decrease of discharge current with the increase of magnetic field. Such contradiction was possible to be explained after discovering the so called "effect of anode alignment" [7,8]. The anode is considered to be aligned, if the uniform magnetic field is parallel to anode surface, the anode is sufficiently long to neglect the end effects and there are no protuberances on the anode, the role of which can be played, in particular, by the inaccuracy of the device construction, by the flaked pieces of sputtered film, and so on. It was experimentally shown that at careful alignment, when the size of misalignment (the height of protuberance, or the value of $\alpha L$, where $L$ is the length of anode, and $\alpha$ is the angle between the anode axis and uniform magnetic field) does not exceed the size of electron

Larmor radius, the discharge current increases with the magnetic field both in the magnetron geometry [9] and in the geometry of inverted magnetron [7,8]. At the same time, any small anode misalignment (protuberance, lack of parallelism, non-uniformity of the magnetic field) causes the saturation of discharge current with the increase of magnetic field. The initial explanation of this phenomenon was that the misalignment results in the decrease of voltage drop on the discharge electron sheath [7]. However, the experimentally observed dependence of discharge current on the degree of anode misalignment was much stronger, and the value of discharge current for the aligned anode was much less than it was predicted theoretically.

The other phenomenon which is beyond the scope of the existing theory is the ejection of electrons from the electron sheath to the end cathodes. In the discharge, practically there always exists the electron current ejected to the end cathodes and its average value makes 50 % of the value of discharge current. It was observed for the first time more than 50 years ago, but before detecting the vortex structures in discharge electron sheath [10], the mechanism of its origination remained unknown. The process of formation of vortex structures and the ejection of electrons to the end cathodes were studied in [10-14] and described in detail in [15]. Periodically, in the electron sheath the diocotron instability is excited leading to the destruction of azimuthal uniformity of electron sheath and to its fragmentation. Then, from the separate fragments the long-lived vortex structure is formed. The whole process from the excitation of diocotron instability to the formation of stable vortex structure proceeds rather quickly for the period of time much less than the time of electron-neutral collisions. On the contrary, the lifetime of solitary vortex structures is rather large, i.e. much more than the time of electron-neutral collisions. Formation of vortex structures, interaction of vortex structures with each other, periodical displacement of vortex structure to the direction of cathode (taking place in magnetron geometry at low pressures [10]) are accompanied with the pulse ejection of electrons from the sheath to the end cathodes. At the same time, the continuous ejection of electrons takes place from the vortex structure and the neighboring area of electron sheath [16].

Thus, the diocotron instability initiates the chain of nonlinear processes leading to the ejection of a part of sheath electrons to the end cathodes and therefore limits the density of electron sheath. This means that the equilibrium density of electron sheath is determined not by the balance between the ionization and the mobility of electrons across the magnetic field, as it was supposed earlier, but by the "critical" electron density, at which the diocotron instability arises. From this it also follows that the density of electron sheath should not depend on the mechanism of electron transport across the magnetic field. These and other statements made the basis for the model of electron sheath considered in Sec. II. Sec. III deals with the testing of the model of electron sheath. A good agreement is obtained between the theoretical and experimental characteristics of electron sheath both, in magnetron geometry and in the geometry of inversed magnetron. In final IV section, the limits of application of the considered model and its relation to the other models of electron sheath are discussed.

## II. MODEL OF ELECTRON SHEATH

Let us consider an annular cylindrical sheath of gas-discharge nonneutral electron plasma located between two coaxial cylindrical electrodes in strong longitudinal magnetic field. Let $r_a$ be the radius of anode, $r_1$ - the radius of sheath boundary from the anode side, $r_0$ - the radius of sheath boundary from the cathode side, and $r_c$ - the radius of cathode. Let us study simultaneously the both cylindrical geometry: magnetron ($r_c < r_a$) and inverted magnetron ($r_c > r_a$), and take into account the experimentally observed effects having the most significant effect on the processes in electron discharge sheath.

First of all, let us show that the density of electrons in the sheath is limited by the value, at which the diocotron instability arises. Fig.1 taken from [13] gives the typical picture of the sequence of physical processes in the geometry of inverted magnetron.

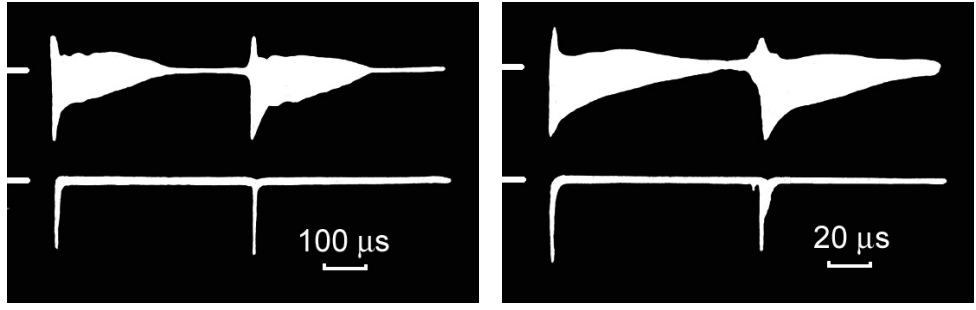

Fig.1. Diocotron instability and vortex structures in inverted magnetron [13]
$r_a = 1.0 cm$; $r_c = 3.2 cm$; $L = 7 cm$; $B = 1.8 kG$; $V = 0.9 kV$; $p = 2\times10^{-6}, 1\times10^{-5} Torr$.

The upper oscillograms show the oscillations of electric field on the anode wall probe. The lower oscillograms show the full current of electrons on the end cathodes. Each pulse of electron current accompanies the development of diocotron instability and the formation of solitary vortex structure. Then follows the quasi-stable state of the sheath, first with slowly "damping" vortex structure, and then, with "quiescent" plasma without structures and oscillations (the left oscillogram). In quasi-stable state the density of sheath electrons increases slowly with time, as is evidenced by the increase of ion current density, measured by means of screened probe located behind the hole in the cylindrical cathode of the inverted magnetron (oscillogram in Fig.2 from [17]). The increase of electron plasma density continues until the density reaches the critical value, at which the diocotron instability is excited. The diocotron instability generates the vortex structures which, in its turn, "initiate" the process of ejection of electrons from the sheath. From the oscillograms given in Figures 2 and 3 one can conclude that: first, the density of sheath with vortex structure is less than the critical density, and second, the diocotron instability arises at the mode, for which the critical density of electrons is minimum.

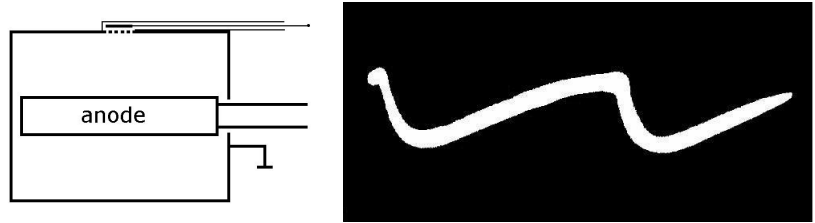

Fig.2. Oscillations of ion current in inverted magnetron [17]
$r_a = 0.9 cm$; $r_c = 4 cm$; $L = 5 cm$; $V = 5 kV$; $B = 1 kG$; $p = 8\times10^{-5} Torr$.

Thus, to determine the maximum density of electron sheath it is necessary to find the critical density of electrons, at which the diocotron instability is excited. For simplicity, let us assume that electron density in the sheath is uniform, and beyond the sheath it equals zero. For such stepwise density profile the conditions of initiation of the diocotron instability were found for the cathode radius equal to zero or to infinity [18]. In [19], the fact that the full charge of discharge electron sheath at large magnetic fields is always equal to anode charge was used, and the conditions of origination of diocotron instability for any cathode radius were obtained. According to [19], let us write the equation for the threshold of diocotron instability in the sheath at $l=1$ mode, as this mode corresponds to the minimum critical electron density

$$r_c^2 \left(r_a^2 + r_0^2 - 2r_0 r_1\right) = r_0^2 \left(r_a^2 - r_1^2\right) \qquad (1)$$

This equation is valid both for the inverted magnetron and the magnetron geometry. In the case of magnetron geometry the condition is imposed on the minimum size of cathode radius (the lower boundary of instability at $l=1$ mode):

$$r_c^2 > \frac{r_0^2\left(r_a^2 - r_1^2\right)}{r_a^2 + r_0^2 + 2r_0 r_1} \tag{2}$$

This condition is fulfilled practically in all experiments with magnetron geometry. It should be note, that if mode $l=1$ cannot or should not be excited for some reason, instead of (1) we should write the condition for the next mode. In such case, all arguments and the following conclusions remain valid, but the electron density in the sheath will be higher.

Equation (1) is one of the main equations of the considered model. The second equation follows from Poisson equation for stepwise distribution of electron density:

$$V_0 = \pi e \left( r_1^2 - r_1^2 \ln \frac{r_1^2}{r_a^2} - r_0^2 + r_0^2 \ln \frac{r_0^2}{r_a^2} \right) n_e \tag{3}$$

Here $V_0$ is the discharge voltage, $n_e$ is the electron density and $e$ is the electron charge.

However, two equations are not enough for determination of three unknowns $n_e$, $r_0$ and $r_1$. As the third equation, let us write the equation for the value of gap between the anode and the sheath. All experiments show that at the large magnetic fields, the electron sheath adjoins the anode surface. Actually, between the anode and the sheath there always exists a small gap and the threshold of the development of diocotron instability is very sensitive to the size of this gap.

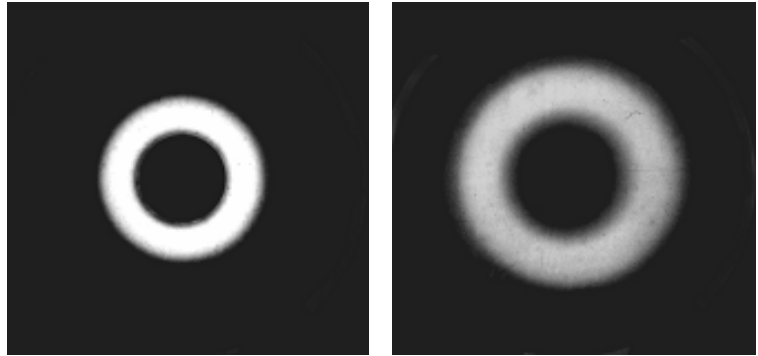

Fig.3. Electron sheath in inverted magnetron at different angles between the anode axis and the magnetic field.
$r_a = 0.9 cm$; $r_c = 3.7 cm$; $L = 7 cm$; $V = 4 kV$; $B = 0.5 kG$; $\alpha = 0$, $\alpha = 0.04 radn$.

Fig.3 shows the photos of electron sheath in inverted magnetron. On the photo on the left the anode is aligned (asymmetry is reduced to the minimum), and on the photo on the right side the angle between the anode axis and the magnetic field equals to $4 \times 10^{-2} radn$. In the first case, the gap between the anode and the sheath is determined by the finiteness of electron Larmor radius, and in the second case, the increase of the gap is caused by the anode misalignment. In the general case, the value of the gap between the sheath and the anode let us present as a sum of anode misalignment and of maximum radial displacement of electron.

$$|r_a - r_1| = d + \Delta r \tag{4}$$

The value of anode misalignment in the case of annular protuberance (ring on the anode) is equal to the height of this protuberance $h$, and in the case of non-parallelism of anode axis to magnetic field it equals to $\alpha L$, where $L$ is the anode length, and $\alpha$ is the angle between the anode axis and the magnetic field

$$d = \begin{cases} h \\ \alpha L \end{cases} \tag{5}$$

Here we make the following assumptions: a slight misalignment at non-parallelism of anode axis to magnetic field, in the first approximation does not cause the geometrical asymmetry of the sheath, it results only in the increase of the gap between the anode and the electron sheath. Indeed, as it is seen from the photo, the electron sheath retains almost a circular shape even at the tilting of magnetic field about the anode axis. Note that here we speak about the influence of misalignment on the threshold of initiation of diocotron instability. The influence of field asymmetry on the processes taking place in the electron sheath is considered below.

The value of radial displacement of electron ($\Delta r$) starting from the radius $r_1$ towards the anode can be obtained by solution of the equations of electron motion in cylindrical geometry:

$$\Delta V = \frac{eB^2}{8mc^2}\left((r_1 \mp \Delta r)^2 - r_1^2\right)^2 (r_1 \mp \Delta r)^{-2}, \tag{6}$$

where $\Delta V$ is the potential drop on $\Delta r$ section, $m$ is the electron mass, $c$ is the velocity of light, $B$ is the magnetic field. Here and below the upper sign corresponds to the inverted magnetron, and the lower sign – to the magnetron geometry. By means of simple manipulations we obtain the third equation of the considered model:

$$B^2\left((r_a \pm d)^2 - r_1^2\right)^2 = 8\pi mc^2 (r_a \pm d)^2 (r_0^2 - r_1^2)\left(\ln r_1^2 - \ln(r_a \pm d)^2\right)n_e \tag{7}$$

Set of equations (1), (3), (7) allows us to determine $n_e$, $r_0$ and $r_1$ for any geometrical ($r_a, r_c, L$) and electrical ($B, V_0$) discharge parameter, as well as for the given value of anode misalignment ($d$). The model is rather simple. It does not contain any empirical coefficients, indeterminate parameters and thus, can be easily used for comparison of experimental and theoretical characteristics of electron sheath. By using dimensionless variables and notations

$$n = \frac{4\pi mc^2}{B^2}n_e \qquad \rho_0 = \frac{r_0}{r_a} \qquad \rho_1 = \frac{r_1}{r_a}$$
$$\varphi_0 = \frac{mc^2}{e}\frac{V_0}{r_a^2 B^2} \qquad \rho_c = \frac{r_c}{r_a} \qquad \delta = \frac{d}{r_a} \tag{8}$$

the set of equations (1), (3), (7) can be written in more compact form:

$$\rho_c^2\left(1 + \rho_0^2 - 2\rho_0\rho_1\right) = \rho_0^2\left(1 - \rho_1^2\right) \tag{9}$$

$$4\varphi_0 = n\left(\rho_1^2\left(1 - \ln\rho_1^2\right) - \rho_0^2\left(1 - \ln\rho_0^2\right)\right) \tag{10}$$

$$2n\left(\rho_0^2 - \rho_1^2\right)\left(\ln\rho_1^2 - \ln(1 \pm \delta)^2\right)(1 \pm \delta)^2 = \left((1 \pm \delta)^2 - \rho_1^2\right)^2 \tag{11}$$

## III. ELECTRON SHEATH MODEL TESTING

The considered model of electron sheath allows us determine the electron density and the geometrical dimensions of the sheath at the moment directly preceding the initiation of diocotron instability. The evolution of diocotron instability, the formation of quasi-stable vortex structure and the ejection of electrons to the end cathodes is the rapid, nonlinear, collisionless process, during which the sheath losses a part of electrons. Then, as a result of ionization, the electron density begins to increase. Thus, the electron density in the sheath periodically changes and for correct comparison of experimental and theoretical results, it is necessary in model to use the time average value of electron density $n_e$, instead of its maximum value $n_{cr}$.

The estimations show that in the inversed magnetron the average value of electron density is in the range of $n_{cr} > n_e > 0.7 n_{cr}$. Therefore, we can assume that $n_e = n_{cr}$ is a rather good approximation for the inverted magnetron and make the quantitative comparison between the theory and the experiment.

In magnetron geometry, the vortex structures in the sheath exist continuously, and consequently, the electron density is always less (possibly significantly less) than the critical value. Therefore, for the magnetron we will make the comparison between the theory and the experiment both for $n_e = n_{cr}$, and $n_e = 0.5 n_{cr}$, in order to follow the tendency to the agreement between the theory and the experiment at the decrease of $n_e$. (In this case the following procedure of calculations is used: first, from (1), (3) and (7) $n_{cr}$ is found, then we take $n_e = 0.5 n_{cr}$, and for this value of $n_e$ find $r_0$ and $r_1$ from (3) and (7)).

Fig.4 shows the dependencies of the thickness of electron sheath $|r_0 - r_1|$ on the magnetic field (solid lines) calculated for the magnetron geometry and the geometry of inverted magnetron. In the same figure the experimental values $|r_0 - r_1|$ are plotted, which are determined from the photos of electron sheath.

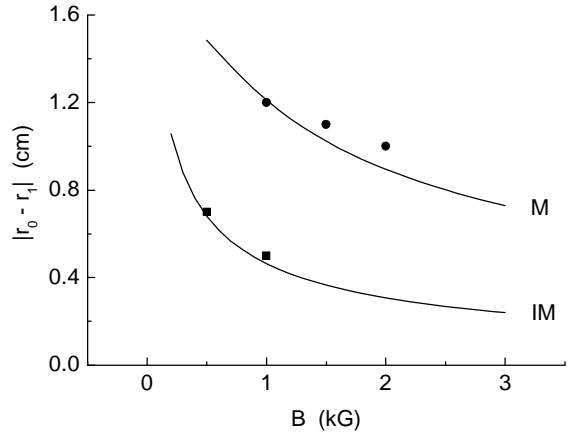

Fig.4. Thickness of electron sheath in magnetron (M) and inverted magnetron (IM)
M: $r_a = 3.2 cm$; $r_c = 0.9 cm$; $L = 7 cm$; $V = 4 kV$; $\alpha = 0$
IM: $r_a = 0.9 cm$; $r_c = 3.2 cm$; $L = 7 cm$; $V = 4 kV$; $\alpha = 0$

In Fig.5, the solid lines show the theoretical dependence of electron sheath density on the angle between the anode axis and the magnetic field in the inverted magnetron. The dots in the figure show the experimental values of electron density from [20].

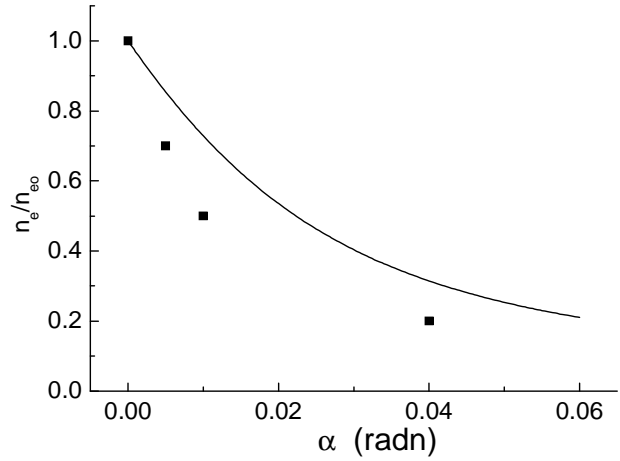

Fig.5. Dependence of electron density in the inverted magnetron on the angle between the anode axis and magnetic field
$r_a = 0.9 cm$; $r_c = 3.2 cm$; $L = 7 cm$; $V = 4 kV$; $B = 1.0 kG$; $p = 8 \times 10^{-5} Torr$.

Fig.6 shows the theoretical dependencies (solid lines) of the anode electric field on the magnetic field for the magnetron and the inverted magnetron. In the same figure, the experimental values of anode electric fields are shown by dots, taken from [21]. As it is seen from the figure, the agreement between the theory and the experiment is better for the inverted magnetron. In case of magnetron geometry, the agreement is improved significantly, if we take $n_e = 0.5 n_{cr}$ (dotted curve).

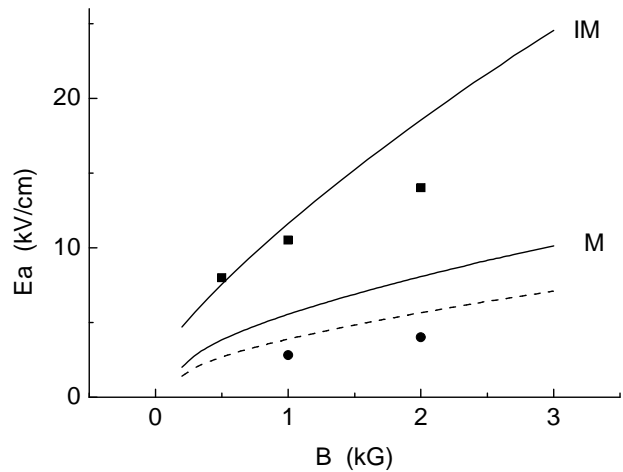

Fig.6. Electric field on the anode in magnetron (M) and inverted magnetron (IM)
IM: $r_a = 1.0 cm$; $r_c = 3.2 cm$; $L = 7 cm$; $V = 4 kV$; $p = 5 \times 10^{-5} Torr$; $\alpha = 0$
M: $r_a = 3.2 cm$; $r_c = 1.0 cm$; $L = 7 cm$; $V = 4 kV$; $p = 5 \times 10^{-5} Torr$; $\alpha = 0$

It should be noted that the number of experimental measurements of the sheath thickness, of the average electron density and of the electric field on the anode surface is insignificant. Therefore, to make more full and accurate quantitative comparison of the theory with the experiment, we use the ion current, as there are sufficient direct measurements of the dependence of the ion current on different discharge parameters. The value of ion current does not depend on the mechanism of electron transport across the magnetic field. Ions are generated in electron sheath at the expense of ionization of neutral gas atoms and leave the discharge gap without collisions. Therefore, the value of ion current is equal to

$$J_i = 2\pi Le \int_{r_0}^{r_1} v_i n_e r dr = \pi Le \overline{v_i} n_e \left(r_0^2 - r_1^2\right) \quad (12)$$

Here $v_i$ is the frequency of ionization of neutral atoms by electrons. The average frequency of ionization ($\overline{v_i}$) in the discharge electron sheath was measured in the geometries of magnetron and inverted magnetron [21]. For the aligned anode it is well described by the following dependence:

$$\overline{v_i} = v_{i0} \ln(1+\varepsilon^2)/2\varepsilon \quad (13)$$

Here $\varepsilon = \lambda E_a / B$, where $E_a$ is the electric field on the anode, $B$ is the magnetic field, $\lambda$ is the coefficient ($\lambda = 0.19(kG \cdot cm/kV)$); and $v_{i0}$ is the maximum frequency of ionization for monochromatic electron beam, which, in case of argon is equal to $v_{i0} = 2\times 10^{-7} n_0$ [22], where $n_0$ is the neutral gas density. In the inverted magnetron one can use the approximation $\overline{v_i} = 0.4 v_{i0}$ [21] with a rather good accuracy.

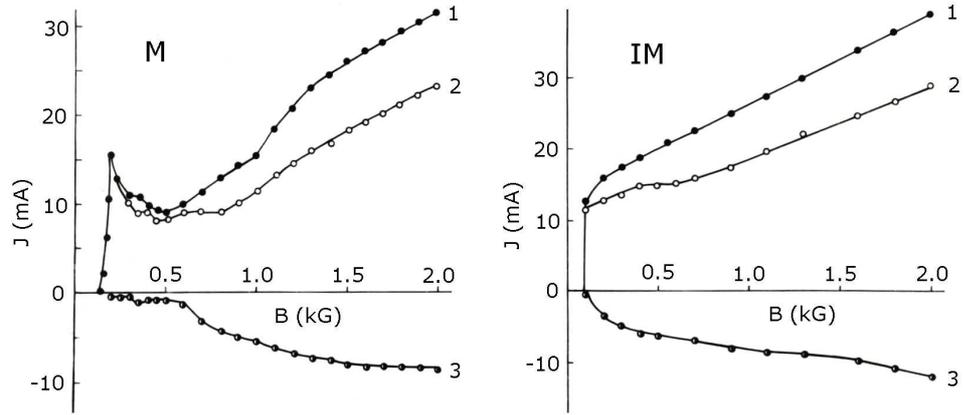

Fig.7. Experimental dependences of ion current (1), discharge current (2), and electron current on the end cathodes (3) versus magnetic field in magnetron (M) and inverted magnetron (IM)

M: $r_a = 3.2 cm$; $r_c = 0.9 cm$; $L = 7 cm$; $V = 4kV$; $p = 1.5 \cdot 10^{-4} Torr$; $\alpha = 0$

IM: $r_a = 0.9 cm$; $r_c = 3.2 cm$; $L = 7 cm$; $V = 4kV$; $p = 3 \cdot 10^{-4} Torr$; $\alpha = 0$

Before starting the process of comparison of experimental and theoretical values of ion current, note that the experimental dependence of discharge current (electron current on the anode) on magnetic field and on other discharge parameters is similar to the dependence of ion current. The difference is that the ion current is 1.3-1.5 times greater than the discharge current.

Fig.7 shows the magnetic field dependencies of ion current on the cylindrical cathode, discharge current on the anode and electron current on the end cathodes in magnetron and inverted magnetron for $\alpha = 0$. As it is seen from the figure, the dependencies for all three currents have the same behavior except for the region of small magnetic fields. Following from this, we can make the comparison of ion current calculated theoretically not only with the experimental values of ion current, but also with the experimental values of discharge current (in the cases, when the measurement of ion current was not made separately).

Fig.8 shows the dependence of ion current $J_i / J_{i0}$ ($J_{i0}$ is the ion current at $\alpha = 0$) on the angle between the anode axis and the magnetic field in the geometry of inverted magnetron. The solid line shows the theoretical dependence (by dotted line the theoretical dependence is shown for approximated values of $\overline{v}_i = 0.4 v_{i0}$ ). In the figure, the experimental values of ion current are shown by dots at the nonuniformity of magnetic field $\Delta B / B = 0.016$, taken from [8], and by circles - at $\Delta B / B = 0.002$. Here, $\Delta B$ is the nonuniformity of magnetic field on the anode length.

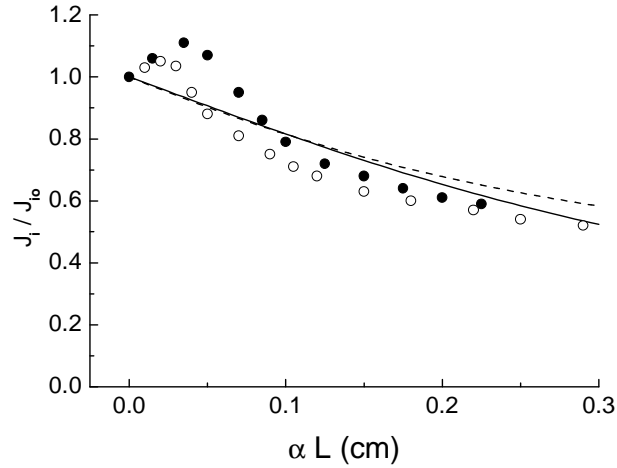

Fig.8. Dependence of ion current in inverted magnetron on the angle between the anode axis and the magnetic field.

• – $r_a = 0.9 cm$; $r_c = 3.0 cm$; $L = 5 cm$; $V = 4 kV$; $B = 1 kG$; $p = 2.7 \times 10^{-4} Torr$; $(\Delta B / B) = 0.016$

○ – $r_a = 0.9 cm$; $r_c = 3.2 cm$; $L = 7 cm$; $V = 4 kV$; $B = 1 kG$; $p = 3 \times 10^{-4} Torr$; $(\Delta B / B) = 0.002$

As it is seen from the Fig.8, the agreement between the experimental and theoretical dependencies is rather good. The hump on the experimental dependence of ion current at small angles $\alpha$ ($\alpha L < 0.1$) can be related with the formation of internal resonance sheath. Fig.9 shows the photo of electron sheath with internal resonance sheath in the inverted magnetron.

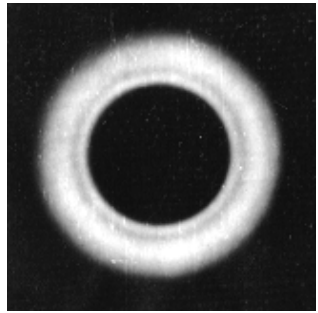

Fig.9. Resonance sheath in inverted magnetron
$r_a = 0.9 cm$; $r_c = 3.7 cm$; $L = 7 cm$; $V = 4 kV$; $B = 0.5 kG$; $\alpha = 0.01 radn$

Resonance sheath was first found in [8] and investigated in detail in [20]. It was shown that at small angles $\alpha$, the resonance takes place between the axial and azimuthal motions of electron and therefore, near the anode surface the resonance sheath is formed in spite of the fact that the magnetic line passing in this region of the sheath intersects the anode surface. At $\alpha L > 0.1$ the longitudinal velocity of electrons increases to such an extent that the resonance is disturbed and the internal sheath disappears (electrons of resonance sheath fall on the anode along the magnetic field lines). As it follows from the model of electron sheath, the decrease of the gap between the anode and the sheath leads to the increase of critical electron density, hence, to the increase of ion current. Therefore, the resonance sheath can be the reason of the appearance of hump on the experimental dependence of ion current on $\alpha$ angle.

Fig.10 shows the dependence of ion current $J_i / J_{i0}$ on angle $\alpha$ between the anode axis and the magnetic field in magnetron geometry. The solid line is for theoretical dependence and the dots show the experimental values of ion current at $\Delta B / B = 0.002$ taken from [9].

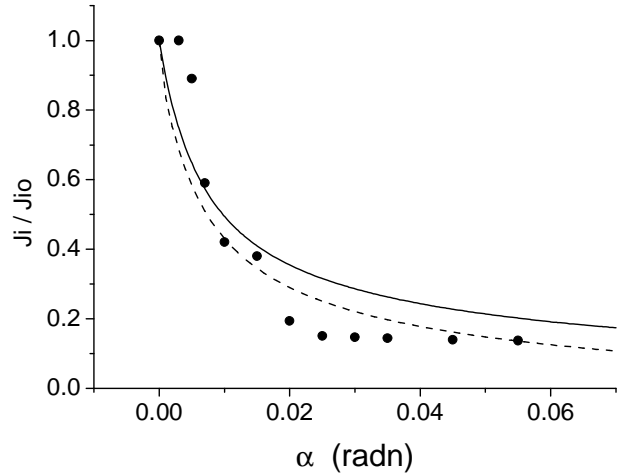

Fig.10. Dependence of ion current in magnetron on the angle between the anode axis and the magnetic field
$r_a = 3.2 cm$; $r_c = 0.9 cm$; $L = 7 cm$; $V = 4 kV$; $B = 1.8 kG$; $p = 1.5 \times 10^{-4} Torr$

As it is seen from the Fig.10, for magnetron geometry the agreement between the theory and the experiment is worse than for the geometry of inverted magnetron. The agreement is improved, if we take $n_e = 0.5 n_{cr}$ (dotted curve). Discrepancy between the theory and the experiment at small angles $\alpha$ can be related, as in the case of the geometry of inverted magnetron, with the formation of resonance electron sheath near the anode surface.

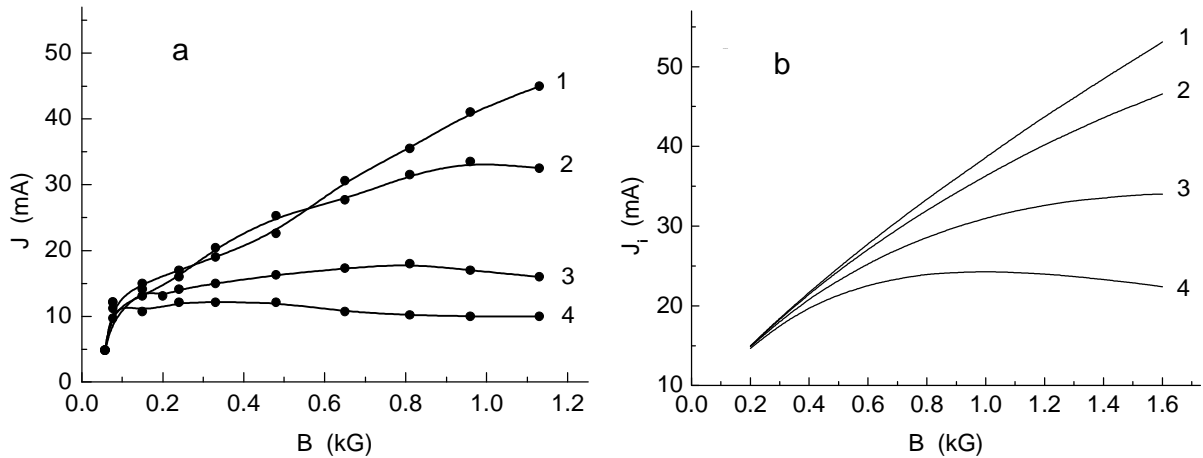

Fig.11. Dependence of ion current in inverted magnetron on the magnetic field at different thicknesses of rings on the anode
$r_a = 0.9 cm$; $r_c = 5.0 cm$; $L = 7 cm$; $V = 4 kV$; $p = 3 \cdot 10^{-4} Torr$; $1 - h = 0$, $2 - 0.03$, $3 - 0.1$, $4 - 0.2 cm$.

Now, let us compare the theoretical and experimental dependencies of ion current on magnetic field at anode alignment and at the anode misalignment. Fig.11a shows the experimental dependencies of discharge current on $B$ in inverted magnetron for different thicknesses ($h$) of small rings, slipped on the anode [7]. In this experiment, the ion current was not measured separately. Therefore, we will use the above-mentioned similarity of experimental dependencies of ion current and discharge current on magnetic field and will compare the experimental curves in Fig.11a with the theoretical dependencies of ion current on magnetic field shown in Fig.11b for the same parameters of discharge and ring thicknesses. In this experiment, the anode misalignment was not accompanied with azimuthal asymmetry of sheath and, as is seen from the figure, the agreement between the experimental and theoretical dependencies is quite good.

Fig.12 shows the experimental (a) and theoretical (b) dependencies of ion current on magnetic field at different angles $\alpha$ between the anode axis and the magnetic field in magnetron geometry of discharge device. The experimental dependencies are taken from [9].

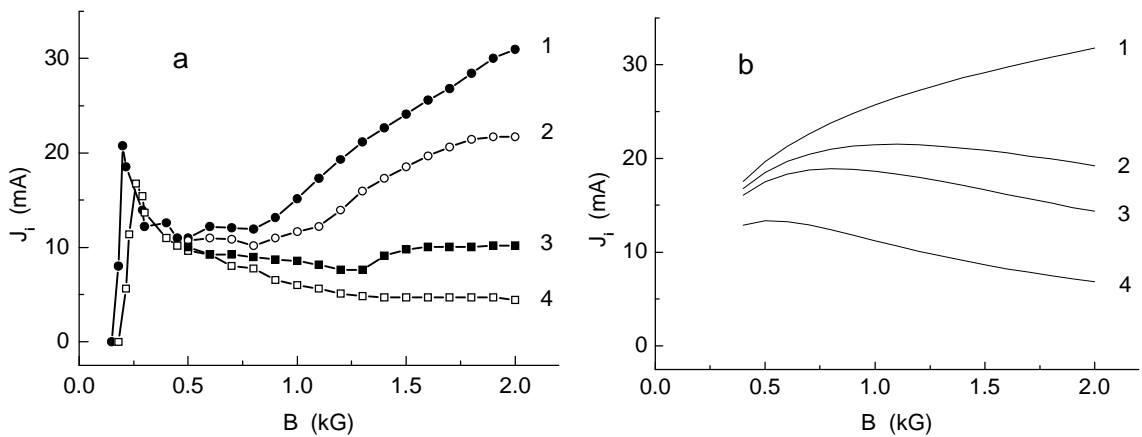

Fig.12. Dependence of ion current in magnetron on magnetic field at different angles between the anode axis and the magnetic field
$r_a = 3.2 cm$; $r_c = 0.9 cm$; $L = 7 cm$; $V = 4 kV$; $p = 1.5 \cdot 10^{-4} Torr$;
$1 - \alpha = 0$, $2 - 0.005$, $3 - 0.01$, $4 - 0.05 radn$.

As we see from Fig.12 the agreement between the theory and the experiment is worse in the case of magnetron geometry than in the case of inverted magnetron with rings. This can be explained by the fact that in theoretical calculations the electron density is taken too high ($n_e = n_{cr}$) and the azimuthal asymmetry of sheath is not considered.

### IV. DISCUSSION AND CONCLUSION

Let us consider some problems connected with the correct use of our model of electron sheath and with the scope of its application. First of all, let us dwell on the problem of alignment, as, at testing of the model one of the main criteria of comparison of the theory with the experiment was the dependence of the characteristics of electron sheath on the value of anode misalignment.

In the experiment three methods of misalignment were used. The first method consists in the following: on the end of preliminarily aligned anode small rings of different thicknesses are slipped on [7]. This method is connected with switching the discharge off when changing the rings and with other inconveniencies, however, it does not cause the azimuthal asymmetry of the sheath and therefore, is the experimental analogue being the most similar to the theoretical model. Just the results of measurements with such misalignment are in good agreement with the theoretical model of electron sheath.

The second method is the change of the angle between the anode axis and the magnetic field [8,9]. It allows us to change smoothly the tilt of magnetic field by turning the solenoid without switching the discharge off. This method makes the controlled azimuthal and axial asymmetries of electric and magnetic fields in the discharge electron sheath and therefore, is especially useful for studying the effects connected with the asymmetry of fields, e.g. with the formation of resonance sheath [20]. Despite the appearance of asymmetry, the results of measurements with such misalignment are in satisfactory agreement with the theoretical model of electron sheath not considering the asymmetry of fields. This gives the evidence of the correctness of the used assumptions at the development of theoretical model of sheath.

The third method consists in the change of uniformity of magnetic field. For estimation of nonuniformity $\Delta B / B$ value was used. Here, $B$ is the value of magnetic field in the central part of solenoid, just where the center of anode was located, and $\Delta B$ is the change of magnetic field along the anode length (the difference in the value of magnetic field between the central part of the anode and its ends). In this case too, the azimuthal symmetry of the sheath was retained. In the first works with inverted magnetron, when the effect of anode alignment was discovered [7,8] the nonuniformity of magnetic field was $\Delta B / B = 0.016$. Later, by means of correcting coils, the nonuniformity of magnetic field was improved up to $\Delta B / B = 0.002$, allowing the observation of the effect of anode alignment in the magnetron geometry [9]. From this it follows that, for obtaining the increasing dependence of discharge current on magnetic field, the uniformity of magnetic field in the magnetron geometry should be much better than in the geometry of inverted magnetron. However, there are no problems specific for the geometry and this effect can be easily explained. In ordinary solenoid the magnetic field is the maximum at the center of solenoid and decreases towards its edges. Fig.13 (left side) shows the approximate shape of magnetic field lines inside the solenoid and the location of anode in case of magnetron (M) and inverted magnetron (IM). The same figure shows the dependence of discharge current on magnetic field in the magnetron for different values of $\Delta B / B$ [9].

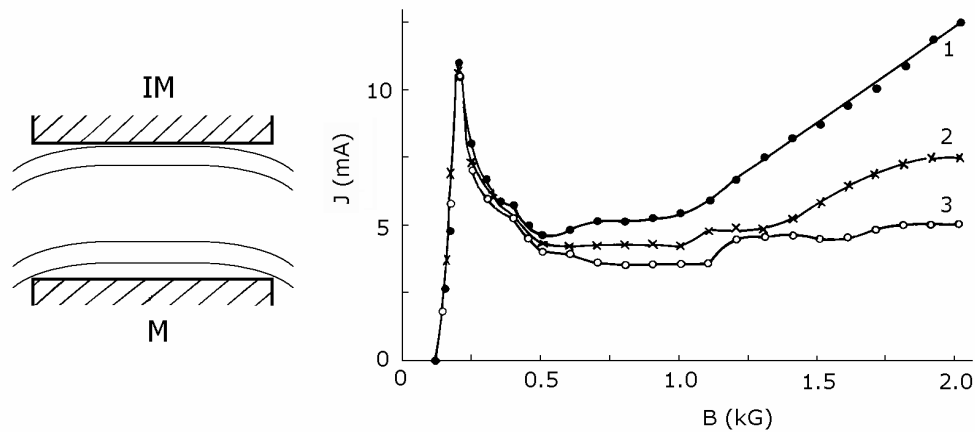

Fig.13 Influence of magnetic field nonuniformity on the dependence of discharge current on magnetic field in magnetron geometry [9].
$r_a = 3.2 cm; r_c = 0.9 cm; L = 7 cm; V = 4 kV; p = 8 \cdot 10^{-5} Torr$;
$1 - (\Delta B / B) = 0.002, 2 - 0.008, 3 - 0.016$

As is evident from Fig.13 in inverted magnetron the nearest line of magnetic field not intersecting the anode adjoins the anode surface almost along its length, while in the magnetron geometry the same magnetic field line along almost the whole anode length is at some distance from its surface. Therefore, for the alignment of anode, the requirements for the uniformity of magnetic field in magnetron geometry or in Penning cell are greater than in inverted magnetron.

Now let us study the scope of applicability of the considered model, and dwell as well on some other models of electron sheath. If in the discharge there are not the diocotron instability, vortices, and electron ejection from the sheath to the end cathodes, instead of equation (1) one should use the equation of continuity for electrons. In this case, the value of electron density is determined from the balance between the ionization and the mobility of electrons across the magnetic field. In case of classical transverse mobility, the density of electrons will be equal to [7,23]

$$n_e = \left(v_i / v_0\right)\left(B^2 / 4\pi mc^2\right), \qquad (14)$$

where, $v_0$ is the total frequency of electron-neutral collisions. This model, often called the diffusion model, was the first model of discharge electron sheath. As it was mentioned above, the diffusion model does not allow the quantitative description of the characteristics of discharge electron sheath at large magnetic fields. It is more suitable for the description of stable electron plasma: a gas-discharge electron sheath at small magnetic fields, when the electron sheath fills the whole discharge gap, and a column of pure electron plasma in Penning-Malmberg cell at the pressures of neutral gas $p > 10^{-7} Torr$ [24].

For ion accelerators with closed electron drift (thrusters), the one-Larmor model of discharge electron sheath was developed, when the density of ions accelerated in the sheath is comparable with the density of electrons, and the whole fall of accelerating voltage takes place on the electron sheath with the thickness of about one Larmor radius [25]. Such sheath can accelerate large ion currents much exceeding the possibilities of ordinary electrostatic accelerators. The schematic model is as follows: the cathode plasma, which is the unlimited source of electrons, abuts on to the sheath from the cathode side, and the anode plasma, which is the source of accelerated ions – from the anode side. The sheath is considered to be collisionless, as the electrons from the cathode plasma enter it, and being reflected from the magnetic field return back to the cathode plasma. Relatively seldom collisions of electrons with neutrals in the sheath leading to their losses are not considered by the theory. The magnetic field in accelerating sheath usually is less than in anode plasma, as it is partially compensated by intrinsic magnetic field of Hall current of the electrons.

In [26] the dynamic model of discharge electron sheath at large magnetic fields is proposed, according to which, the electron sheath is periodically formed near the cathode and then drifts to the anode. The velocity of sheath motion is determined by the classical electron mobility across the magnetic field. In the sheath a continuous process of ionization takes place and the excessive electrons from the external boundary of sheath go to the end cathodes along the magnetic field. Approaching the anode, the sheath is absorbed by the anode. Then, the electric field appears at the cathode and the cycle is repeated. In this model, the experimental facts taking place in the discharge at large magnetic fields are neglected: namely, the sheath is located near the anode surface, in the sheath the diocotron instability is excited periodically, and a vortex structures are formed [10-13]. If the sheath is periodically displaced from cathode to anode, as the authors of the model suppose, not only the region near the anode (Fig.3), but the whole discharge gap would glow, the ejection of electrons to the end cathodes not only from the region near anode [7,8,20], where the sheath is located, but from the whole discharge gap would take place, and finally, the duration of the ejection of electrons would be, at least, of the order of several collision time, but not much less than the time of electron-neutral collisions, as it is observed by the experiment.

Let us return to our model of electron sheath. This model is applicable at large magnetic fields, when the sheath is not in contact with the cathode. If we begin to decrease the magnetic field, the sheath will begin to extend, and when it reaches the cathode surface, the diocotron instability cannot be excited any more, and the vortices and the ejection of electrons disappear together with it. Thus, the transition to the diffusion model of electron sheath will take place. If we judge from Fig.7, in the geometry of inverted magnetron our model is applicable almost until the discharge ignition, but in magnetron geometry, the transition to the diffusion model should take place at the magnetic field $B_1 \approx 0.5 kG$. At the transition to the diffusion model, electron Larmor diameter near the cathode surface is practically equal to zero. At the further decrease of magnetic field, it will increase until it reaches the dimensions of discharge gap. Thus, we will go to the one-Larmor regime of electron sheath. In thrusters, the cathode plasma serves as a source of filling the one-Larmor sheath with electrons. In our case, the cathode plasma is absent, however, in the case of magnetron geometry its role is played by electron-electron emission from the cathode surface. This effect was first discovered in multicavity magnetrons. Then it was shown that the similar effect takes place in the magnetron geometry of discharge device [27]. Fig.14 from [28] gives the magnetic field dependencies of ion and electron currents on the cylindrical cathode in magnetron geometry of discharge device. The probes for measuring these currents were placed inside the cylindrical cathode, and the separation of ions and electrons was made by magnetic field. The electrons were registered with the energy of not less than 50 eV.

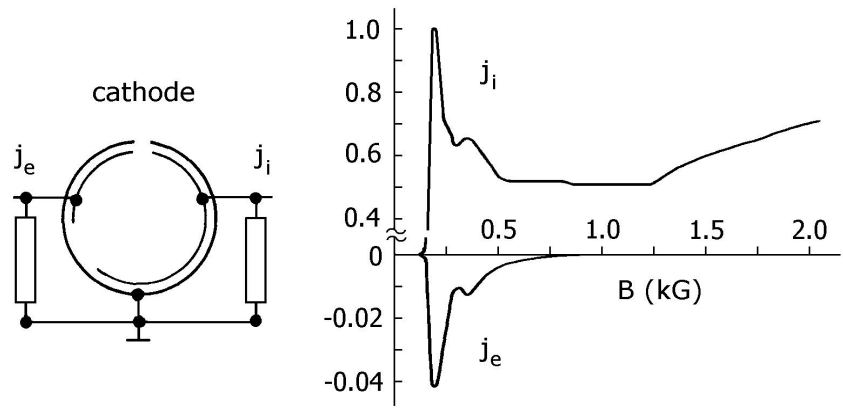

Fig.14. Inverse bombardment of cathode with electrons in magnetron geometry [28]
$r_a = 3.2 cm$; $r_c = 0.75 cm$; $L = 7 cm$; $V = 4 kV$; $p = 2 \cdot 10^{-4} Torr$

From Fig.14 It can be seen that immediately after the ignition of discharge, the intense electron bombardment of cathode takes place with the electrons of sufficiently high energy, and the maximum current of these electrons coincides with the peak of discharge current at the magnetic field equal to about $B_0 = 0.2 kG$. The bombardment of cathode with the electrons of high energy extends to the region of magnetic fields up to $0.7 kG$.

Now, let us start the estimation of discharge currents at magnetic fields $B = B_1$ and $B = B_0$. In both cases, the sheath occupies practically almost the whole volume of discharge gap. In the first case, we have the diffusion sheath and the discharge current equals the ion current $J_1 = J_i = \pi e L (r_a^2 - r_c^2) n_e v_i$, where the electron density depends on the discharge voltage. In the second case we have one-Larmor electron sheath, and any collision of electron with neutral leads to its ejection to the anode. Therefore, neglecting the ionization, the discharge current equals the current of electrons undergoing the collision with neutrals $J_0 = J_e = \pi e L (r_a^2 - r_c^2) n_e v_0$. We do not know the value of electron density for one-Larmor regime with secondary emission cathode. It depends on emission properties of cathode. But for estimation, let us take the maximum value equal to the electron density of diffusion sheath. Then the current ratio will be equal to $J_0/J_1 = v_0/v_i \approx 10$. The experimental value of this ratio is less and depends on the experimental conditions. For the dependence of discharge current on magnetic field, shown in Fig.7, it is equal to 2 and for the dependence shown in Fig.13 – to 2.4.

In conclusion let us note that the model of discharge electron sheath considered here is based on diocotron instability, though the main role in limitation of electron density is played by the vortex structures appearing in this case and by their interaction with the electron sheath. In the model we have made several assumptions: the step-function density profile, keeping the cylindrical form of the sheath at the tilting of magnetic field, quasi-stability and some other assumptions. However, in spite of all simplifications, the model gives quantitative and qualitative description of electron sheath characteristics, not depending on the mechanism of electron transport across the magnetic field, and allows for the first time the quantitative description of the influence of anode misalignment on the characteristics of discharge electron sheath.